\documentclass[12pt]{article}

\usepackage{amssymb,amsmath}

\usepackage{graphicx}
\DeclareGraphicsExtensions{.pdf,.png,.jpg}

\renewcommand{\theequation}{\arabic{section}.\arabic{equation}}
 \catcode`\@=11 \@addtoreset{equation}{section}\catcode`\@=12
\newcommand{\R}{ {\mathbb R} }

\newcommand{\fnm}{\footnotemark}
\newcommand{\fnt}{\footnotetext}

\begin{document}

 \begin{center}

 \large \bf 
 Stable  exponential cosmological solutions with 
 two factor spaces in the Einstein-Gauss-Bonnet  model  with a $\Lambda$-term
  
  \end{center}

 \vspace{0.3truecm}

\begin{center}

 V. D. Ivashchuk$^{1,2}$ and A. A. Kobtsev$^{3}$ 

\vspace{0.3truecm}

 \it $^{1}$ Institute of Gravitation and Cosmology,
   Peoples' Friendship University of Russia (RUDN University),
   6 Miklukho-Maklaya St.,  Moscow 117198, Russian Federation \\ 

 \it $^{2}$ Center for Gravitation and Fundamental Metrology,
 VNIIMS, 46 Ozyornaya ul., Moscow 119361,  Russian Federation
 
 \it $^{3}$ Institute for Nuclear Research of the Russian Academy of Sciences,
            Moscow, Troitsk, 142190,  Russian Federation

\end{center}

\begin{abstract}

We study   $D$-dimensional Einstein-Gauss-Bonnet  gravitational model including  the Gauss-Bonnet term and the cosmological term $\Lambda$. We find a class of solutions with  exponential time dependence of two scale factors, governed by two Hubble-like parameters $H >0$ and $h$, corresponding to factor spaces of dimensions $m >2$ and $l > 2$, respectively. These solutions contain
a fine-tuned $\Lambda = \Lambda (x, m, l, \alpha)$, which depends upon the ratio $h/H = x$, dimensions 
of factor spaces $m$ and $l$, and the ratio  $\alpha = \alpha_2/\alpha_1$  
of two constants ($\alpha_2$ and $\alpha_1$) of the model.  
The master equation $\Lambda(x, m, l,\alpha) = \Lambda$
is equivalent to a polynomial equation of either  fourth or third order and may be solved
in radicals. The explicit solution for $m = l$ is presented in Appendix. Imposing certain restrictions on $x$,  we  prove  the stability of the solutions  in a class of cosmological solutions with diagonal  metrics. 
We also consider a subclass of solutions with small enough variation of the effective gravitational constant $G$ and show the stability of all solutions from this subclass.

\end{abstract}


\newpage

\section{Introduction}

In this paper  we deal with the so-called  Einstein-Gauss-Bonnet (EGB) gravitational model 
in $D$ dimensions, which contains  Gauss-Bonnet term and cosmological term $\Lambda$.  
The so-called Gauss-Bonnet term appeared in (super)string theory  as a correction 
(in appropriate frame) to the  effective (super)string action 
(e.g. heterotic one)  \cite{Zwiebach}-\cite{MTs}. 

It should be noted that at present EGB gravitational model, e.g. with 
cosmological term $\Lambda$, and  its modifications, 
see   \cite{Ishihara}-\cite{Pavl-p-18} and refs. therein,  are under intensive studies in  cosmology,
aimed at  possible explanation  of  accelerating  expansion of the Universe,
which follows from supernovae (type Ia) observational data \cite{Riess,Perl,Kowalski}.

In this paper  we  study the cosmological model with diagonal metric  governed by $n >3$ scale factors
($D = n+1$) depending upon one variable , which is the synchronous time variable. 
As it is well-known,  the multidimensional gravitational model  ($D > 4$) with quadratic in curvatures 
terms governed by the Lagrangian 
$R - 2 \Lambda + \alpha R^2 + \beta  R_{mn} R^{mn} + \gamma R_{mnpq} R^{mnpq}$
leads us for generic values of coupling constants  $\alpha$, $\beta$ and  $\gamma$
  to the fourth-order differential
equations for the components of the metric, while for the unique choice of the constants $\alpha = \gamma = -4 \beta$ 
(yielding  the Gauss-Bonnet term) the equations of motion are of the second order \cite{GG}.  
In cosmological model with the Gauss-Bonnet term the equations of motion are governed by an effective 
Lagrangian which contains $2$-metric (or minisupermetric) $G_{ij}$ and finslerian metric $G_{ijkl}$, see
\cite{Iv-09,Iv-10}  for $\Lambda = 0$ and   \cite{ErIvKob-16, IvKob-18mm} for $\Lambda \neq 0$.  

 Here we consider the cosmological solutions  with exponential dependence of scale factors and obtain 
a class of  solutions with  two scale factors, governed by two Hubble-like parameters $H >0$ and $h$, which correspond to factor spaces of dimensions $m > 2$ and $l > 2$ , respectively, ($D = 1 + m + l$)
and obey relations: $m H + lh \neq 0$ and  $H \neq h$. The solutions depend upon $\Lambda$ and
 $\alpha = \alpha_2/\alpha_1$. The latter is the ratio of two constants of the model: $\alpha_2$ and $\alpha_1$.
 (In string inspired models $\alpha$ corresponds to Regge slope parameter $\alpha'$ which is inverse proportional 
 to the  tension of the string.)
Any of  these solutions  describes   an exponential (e.g. accelerated) expansion  of $3$-dimensional subspace with Hubble parameter $H > 0$ \cite{Ade}.  Here, as in our previous paper \cite{Ivas-16}, we use the Chirkov-Pavluchenko-Toporensky scheme of reduction  (of the set of polynomial equations)  \cite{ChPavTop1}, which gives us a drastical simplification of  the search and analysis of the exact solutions under consideration.  

The special $m =3$ case   was studied recently in ref. \cite{IvKob-m3-18}.
The analysis of the number of solutions for given $\Lambda$ and  $\alpha$ 
(denoted as $n(\Lambda,\alpha) $), carried out in  ref. \cite{IvKob-m3-18}, 
tells us  that  exponential solutions exist if and only the following bounds
 on  $\Lambda$ and $\alpha$   are valid:
\begin{equation}
 \Lambda \alpha \leq  \frac{3l^2 -7l + 6}{8l(l-1)} = \frac{3D^2-31D+82}{8(D - 4)(D - 5)} = \lambda_c
 \label{0.i1}
\end{equation}
 for $\alpha > 0 $ and 
\begin{equation}
 \Lambda |\alpha| > \frac{(l+2)(l+3)}{8l(l +1)} = \frac{(D -2)(D-1)}{8(D - 4)(D - 3)} =|\lambda_a|
 \label{0.i2}
\end{equation}
for $\alpha < 0$. Here $D = 4 + l$.

Relations (\ref{0.i1}) and (\ref{0.i2}) were obtained (independently) in ref. 
\cite{Pavl-p-18} \fnm[1]\fnt[1]{ The second  relation (\ref{0.i2}) was extended 
in ref. \cite{Pavl-p-18} to $\Lambda |\alpha| \geq  |\lambda_a|$ by adding into 
consideration the case $H=h$  \cite{ChPavTop,Ivas-16}. In ref. \cite{Pavl-p-18} the cosmological constant $\Lambda_P$  is related to our one as  $\Lambda_P = 2 \Lambda$ and the internal space dimension $l$ is denoted as $D$.}, 
 where they were compared with some other relations obtained in physical literature, e.g. in refs.
\cite{BEMPSS-10} ($\Lambda < 0$, AdS/CFT correspondence), \cite{CG-10} ($\Lambda < 0$, black holes) etc.
In this paper we generalize these bounds to $m \geq 3$ case. 

We  study   the stability of the  solutions in a class of cosmological solutions with diagonal metrics by using results of  refs. \cite{ErIvKob-16,Ivas-16} (see also approach of ref. \cite{Pavl-15}) and show  that the solutions, considered here, are stable if certain restrictions on ratio $h/H$ are imposed.  

Here we also study solutions with a small enough  variation  of the effective gravitational constant 
$G$  in Jordan  frame \cite{RZ-98,I-96} (see also  \cite{BIM,Mel} and refs. therein)
which obey the most severe restrictions on  variation of $G$ from ref. \cite{Pitjeva}. These
solutions are shown to be stable.

We note that a class of stable  solutions with zero variation  of the effective gravitational constant $G$ (either in Jordan or Einstein frames) was considered recently in \cite{ErIv-17-1} 
for $(m,l) \neq  (6,6), (7,4), (9,3)$. Two  special solutions  for $D = 22, 28$ and $\Lambda = 0$
were found earlier in ref.  \cite{IvKob}; in ref. \cite{ErIvKob-16} it was proved that they are stable.
Another six stable  exponential solutions (five in dimensions $D = 7, 8, 9, 13$ and two - for $D = 14$) 
were found recently in \cite{Ivas-16-2}. 

It should be stressed here that  the main physical motivation for this paper and numerous other articles   devoted to higher dimensional cosmological solutions in EGB gravity is based on (i) a possible solution of ``dark energy'' problem, i.e.   explaining  the  accelerated expansion of the Universe. Here, as in some  previous papers, we try to strengthen the physical counterpart of the investigations (e.g.  item (i)) by studying also: (ii) the stability of the obtained solutions and  (iii) by considering the solutions with small enough variation of the effective gravitational constant $G$ (in Jordan  frame) which satisfy the observational restrictions.  At the moment the item (ii) is rarely  considered  in physical publications on multidimensional EGB cosmology while the item (iii) is mostly studied (to our knowledge) in our papers. 

\section{The cosmological model}

The action of the model reads as follows
\begin{equation}
  S =  \int_{M} d^{D}z \sqrt{|g|} \{ \alpha_1 (R[g] - 2 \Lambda) +
              \alpha_2 {\cal L}_2[g] \},
 \label{r1}
\end{equation}
where $g = g_{MN} dz^{M} \otimes dz^{N}$ is the metric defined on
the manifold $M$, ${\dim M} = D$, $|g| = |\det (g_{MN})|$, $\Lambda$ is
the cosmological term, $R[g]$ is scalar curvature,
$${\cal L}_2[g] = R_{MNPQ} R^{MNPQ} - 4 R_{MN} R^{MN} +R^2$$
is the  Gauss-Bonnet term and  $\alpha_1$, $\alpha_2$ are
nonzero constants.

We consider the following (product) manifold
\begin{equation}
   M = \R  \times   M_1 \times \ldots \times M_n 
   \label{r2.1}
\end{equation}
with the cosmological metric
\begin{equation}
   g= - d t \otimes d t  +
      \sum_{i=1}^{n} B_i e^{2v^i t} dy^i \otimes dy^i,
  \label{r2.2}
\end{equation}
where   $B_i > 0$ are arbitrary constants, $i = 1, \dots, n$. In (\ref{r2.1}) 
$M_1, \dots,  M_n$  are one-dimensional manifolds, either  $\R$ or $S^1$, 
and $n > 3$.

The equations of motion for the action (\ref{r1}) 
leads us the to the set of following polynomial equations \cite{ErIvKob-16,IvKob-18mm}
\begin{eqnarray}
  G_{ij} v^i v^j + 2 \Lambda
  - \alpha   G_{ijkl} v^i v^j v^k v^l = 0,  \label{r2.3} \\
    \left[ 2   G_{ij} v^j
    - \frac{4}{3} \alpha  G_{ijkl}  v^j v^k v^l \right] \sum_{k =1}^n v^k 
    - \frac{2}{3}   G_{sj} v^s v^j  +  \frac{8}{3} \Lambda = 0, 
                                                           \label{r2.4}
\end{eqnarray}
$i = 1,\ldots, n$, where  $\alpha = \alpha_2/\alpha_1$. 
Here 
\begin{equation}
G_{ij} = \delta_{ij} -1, 
\qquad   G_{ijkl}  = G_{ij} G_{ik} G_{il} G_{jk} G_{jl} G_{kl}
\label{r2.4G}
\end{equation}
are, respectively, the components of two  metrics on  $\R^{n}$ \cite{Iv-09,Iv-10}: 
2-metric and  Finslerian 4-metric, respectively.
For the case $n > 3$  (or $D > 4$) we have a set of forth-order polynomial  equations.

For $\Lambda =0$  and $n > 3$ the set of equations (\ref{r2.3})
 and (\ref{r2.4}) has  an isotropic
 solution $v^1 = \ldots = v^n = H$ only if $\alpha  < 0$ \cite{Iv-09,Iv-10}.
In \cite{ChPavTop} an isotropic solution was obtained for $\Lambda \neq 0$.

It was proved  in ref. \cite{Iv-09,Iv-10} that there are no more than
three different  numbers among  $v^1,\dots ,v^n$, when $\Lambda =0$. 
This is also valid for the case $\Lambda \neq 0$ when $\sum_{i = 1}^{n} v^i \neq 0$  \cite{Ivas-16}.

\section{Solutions with two Hubble-like parameters}

In this section we find a class of solutions to the set of equations (\ref{r2.3}), 
(\ref{r2.4}) with the following set of Hubble-like parameters
\begin{equation}
  \label{r3.1}
   v =(\underbrace{H,H,H}_{``our" \ space},\underbrace{\overbrace{H, \ldots, H}^{m-3}, 
   \overbrace{h, \ldots, h}^{l}}_{internal \ space}).
\end{equation}
where $H$ is the Hubble-like parameter corresponding  
to an $m$-dimensional factor space with $m > 2$, while  $h$ is the Hubble-like parameter 
corresponding to the $l$-dimensional factor space, $l > 2$. For future 
cosmological applications we split the $m$-dimensional  
factor space into the  product of two subspaces of dimensions $3$ and $m-3$, respectively. 
The first one is considered as ``our" $3d$ space while the second one is identified with 
a subspace of $(m-3 +l)$-dimensional internal space.
 
For a description of an accelerated expansion of a
$3$-dimensional subspace (which may describe our Universe) we put 
\begin{equation}
  \label{r3.2a}
   H > 0. 
\end{equation}

It is  well-known that the (four-dimensional) effective gravitational constant $G = G_{eff}$ in the Brans-Dicke-Jordan (or simply Jordan) frame \cite{RZ-98,I-96}
is proportional to the inverse volume scale factor
of the internal space, see  \cite{BIM} and references therein.

 We note, that due to  ansatz (\ref{r3.1}) ``our'' 3d space expands (isotropically) with 
Hubble parameter $H$ and $(m -3)$-dimensional part of internal space expands (isotropically) with the same Hubble parameter $H$ too. Here  we consider for physical applications (in our epoch) the internal space to be compact one, i.e. we put in (\ref{r2.1}) $M_4 = \ldots = M_n = S^1$). We also put the internal scale factors corresponding to present time  to be small enough as compared to the scale factor of ``our'' space.                    

Due to  ansatz (\ref{r3.1}),  the $m$-dimensional factor space is expanding with the Hubble parameter $H >0$, while  the evolution of the $l$-dimensional factor space is described  by the Hubble-like  parameter $h$.

It was shown earlier in refs. \cite{ChPavTop1,Ivas-16} 
 that the ansatz  (\ref{r3.1}) with two  restrictions on parameters $H$ and $h$  
  imposed 
   \begin{equation}
   m H + lh \neq 0, \qquad  H \neq h,
   \label{r3.3}
   \end{equation}
 leads us to a reduction of the relations (\ref{r2.3}) and (\ref{r2.4})  to the  following set of two polynomial equations
  \begin{eqnarray}
  E =  m H^2 + l h^2 - (mH + lh)^2  + 2 \Lambda 
        - \alpha [m (m-1) (m-2) (m - 3) H^4
          \nonumber \\
       + 4 m (m-1) (m-2) l H^3 h   
       + 6 m (m-1) l (l - 1) H^2 h^2
         \nonumber \\
       + 4 m l (l - 1) (l - 2) H h^3  
       + l (l - 1) (l - 2) (l - 3) h^4] = 0, \quad
         \label{r3.4}   \\
  Q =  (m - 1)(m - 2)H^2 + 2 (m - 1)(l - 1) H h 
        \nonumber \\ 
       + (l - 1)(l - 2)h^2 = - \frac{1}{2 \alpha}.
     \label{r3.5}
  \end{eqnarray}

Using equation  (\ref{r3.5}) we get for $m > 2$ and $l > 2$ 
\begin{equation}
H   =     (- 2 \alpha {\cal P})^{-1/2}, 
 \label{r3.6}
\end{equation}
where we denote
\begin{eqnarray}
{\cal P}  =  {\cal P}(x,m,l)  =   (m - 1)(m - 2) 
  \nonumber \\  
  + 2 (m - 1)(l - 1) x  + (l - 1)(l - 2)x^2, 
 \label{r3.7}  \\
    x  = h/H,
    \label{r3.7x}
 \end{eqnarray}
and 
 \begin{equation}
\alpha {\cal P} < 0. 
 \label{r3.8} 
\end{equation}
We note that the following ``duality'' identity
is obeyed 
\begin{equation}
 {\cal P}(x,m,l) = x^2 {\cal P}(1/x,l,m), 
    \label{r3.8dP}
 \end{equation}
for $x \neq 0$.

Due to restrictions (\ref{r3.3})  we have 
\begin{equation}
  x \neq x_d = x_d(m,l) \equiv  - m/l, \qquad  x \neq x_a \equiv 1, 
 \label{r3.8da} 
\end{equation}
where 
 \begin{equation}
   x_d(m,l) =  1/x_d(l,m).
  \label{r3.8d} 
 \end{equation}
 
 The relation  (\ref{r3.7}) is valid only if        
\begin{equation}
  {\cal P}(x,m,l) \neq 0. 
 \label{r3.8b}
\end{equation}
We note, that for ${\cal P}(x,m,l) = 0$ the relation (\ref{r3.5}) is not obeyed.

The substitution of relation (\ref{r3.6}) into (\ref{r3.4})
leads us to following formulas
\begin{eqnarray}
    \Lambda \alpha = \lambda  =  \lambda(x,m,l) \equiv 
     \frac{1}{4} ({\cal P}(x,m,l))^{-1} {\cal M}(x,m,l)
     \qquad \nonumber \\
      + \frac{1}{8 }( {\cal P}(x,m,l))^{-2} {\cal R}(x,m,l),  
              \qquad    \label{r3.8L}  \\
   {\cal M}(x,m,l) \equiv  m  + l x^2 -(m  + l x)^2, 
              \qquad   \label{r3.8M}  \\
   {\cal R}(x,m,l) \equiv  m (m-1) (m-2) (m - 3)
                 + 4 m (m-1) (m-2) l x 
                \qquad  \nonumber \\    
       + 6 m (m-1) l (l - 1) x^2        
       + 4 m l (l - 1) (l - 2)  x^3
        \qquad  \nonumber \\
       + l (l - 1) (l - 2)(l - 3) x^4.
               \qquad   \label{r3.8R}  
\end{eqnarray}

In what follows we use the following duality identities
\begin{eqnarray}   
   {\cal M}(x,m,l) = x^2{\cal M}(1/x,l,m), 
              \qquad   \label{r3.9dM}  \\
   {\cal R}(x,m,l) =  x^4{\cal R}(1/x,l,m)
                  \qquad   \label{r3.9dR}  
\end{eqnarray}
 for $x \neq 0$.  
 The identities  (\ref{r3.8dP}), (\ref{r3.9dM})  and (\ref{r3.9dR}) imply
 \begin{equation}   
    \lambda(x,m,l) = \lambda (1/x,l,m) 
               \qquad   \label{r3.9l}   
 \end{equation}
for $x \neq 0$.

Using (\ref{r3.8b}) we obtain
 \begin{eqnarray} 
  x \neq x_{\pm} = x_{\pm}(m,l) 
  \equiv \frac{-(m - 1)(l - 1) \pm \sqrt{\Delta(m,l)}}{(l - 1)(l - 2)}, 
 \label{r3.9} \qquad \\
   \Delta(m,l) \equiv  (m - 1)(l - 1)(m + l - 3) = \Delta(l,m).
            \qquad \label{r3.9D}
 \end{eqnarray}
Here $x_{\pm}(m,l)$ are roots of the quadratic equation ${\cal P}(x,m,l) =0$.
These roots obey the following relations
\begin{eqnarray} 
   x_{+}(m,l) x_{-}(m,l) =  \frac{(m - 1)(m - 2)}{(l - 1)(l - 2)}, 
         \qquad \label{r3.10a}  \\
   x_{+}(m,l) + x_{-}(m,l) =  - 2 \frac{(m - 1)}{l - 2}, 
          \qquad \label{r3.10b}
 \end{eqnarray}
which lead us to the  inequalities
\begin{equation} 
    x_{-}(m,l) < x_{+}(m,l) < 0. 
         \qquad \label{r3.11} 
  \end{equation}
These relations and the duality identity (\ref{r3.8dP}) imply 
\begin{equation} 
    x_{\mp}(m,l) = 1/x_{\pm}(l,m) 
         \qquad \label{r3.12} 
  \end{equation}
 for all $m > 2$ and $l > 2$.

Using (\ref{r3.8}) and (\ref{r3.8L}) we get 
 \begin{equation}
   \Lambda   = \alpha^{-1} \lambda(x,m,l),   
               \qquad \label{r3.13a} 
 \end{equation}
where 
\begin{equation} 
    x_{-}(m,l) < x <  x_{+}(m,l) \ {\rm for} \  \alpha > 0  
               \qquad \label{r3.13b}
 \end{equation}    
  and    
   \begin{equation}  
     x <  x_{-}(m,l), \ {\rm or} \ x > x_{+}(m,l) \ {\rm for} \  \alpha < 0.
                 \qquad \label{r3.13c}
 \end{equation}

For $ \alpha < 0$ we have the following limit 
  \begin{equation}  
  \lim_{x \to \pm \infty} \lambda(x,m,l) = 
   \lambda_{\infty}(l) \equiv  - \frac{l(l + 1)}{8 (l - 1)(l - 2)} < 0.
                  \qquad \label{r3.13.l}
  \end{equation}
  Hence 
 \begin{equation}  
   \lim_{x \to \pm \infty} \Lambda = 
   \Lambda_{\infty} \equiv  -  \frac{l(l + 1)}{8 \alpha (l - 1)(l - 2)} > 0,
                  \qquad \label{r3.13.lim.x}
  \end{equation}
$l > 2$. We note that $\Lambda_{\infty}$ does not depend upon $m$.
For $x = 0$ we get in agreement with the duality identity (\ref{r3.9l}) 
\begin{equation}  
   \Lambda = \Lambda_{0} = \alpha^{-1} \lambda(0,m,l) = 
    -  \frac{m(m + 1)}{8 \alpha (m - 1)(m - 2)} > 0,
                  \qquad \label{r3.13.L}
  \end{equation}
$m > 2$. We see that $\Lambda_{0}$ does not depend upon $l$. For $x = 0$
the Hubble-like parameters read 
\begin{equation}
  H  = H_0 = (- 2 \alpha (m - 1)(m - 2))^{-1/2}, \qquad h = 0  
 \label{r3.13.H.0}
 \end{equation}  
and due to relations  (\ref{r2.1}),  (\ref{r2.2}) we get the product
of (a part of) $(m+1)$-dimensional de-Sitter space and $l$-dimensional 
Euclidean space.

{\bf ``Master'' equation.} We rewrite eq. (\ref{r3.8L})  
in the following form
\begin{equation}
       2 {\cal P}(x,m,l) {\cal M}(x,m,l)
       +  {\cal R}(x,m,l) - 8 \lambda ( {\cal P}(x,m,l))^{2} = 0.  
              \qquad    \label{r3.13.M} 
\end{equation}                
This equation may be called as a master equation, since the 
solutions under consideration are governed by it.
The master equation is of fourth order in $x$  
for $\lambda \neq \lambda_{\infty}(l)$ or less (of third order for
$\lambda = \lambda_{\infty}(l)$). 
For any $m > 2$ and $l > 2$ the  equation (\ref{r3.13.M}) may be solved  in radicals, 
though the general  solution has a rather cumbersome form and will not be presented here.
It is worth for any given $m$ and $l$ to find the solution just 
by using Maple or Mathematica. An example of explicit (generic) solution for $m = l > 2$ 
is presented in Appendix A (for  $m = l = 3$ see ref. \cite{IvKob-m3-18}). 
Several special solutions with $m=3$ and $l =4$  were given in ref.  \cite{ErIvKob-16}. 

Now we consider the behaviour of the function  $\lambda(x,m,l)$   in the vicinity of 
 the points $x_{-}(m,l)$ and  $x_{+}(m,l)$. Here the following proposition takes place.

{\bf Proposition 1.} {\em For $m > 2$, $l > 2$ 
 \begin{equation}
    \lambda(x,m,l) \sim B_{\pm}(m,l) (x - x_{\pm}(m,l))^{-2}, 
 \label{r3.13.H}
\end{equation}
 as $x \to x_{\pm} = x_{\pm}(m,l)$, where $B_{\pm}(m,l) < 0$ and hence
 \begin{equation}
   \lim_{x \to x_{\pm}}  \lambda(x,m,l) = - \infty. 
  \label{r3.13.lim}
 \end{equation}
 }

The  Proposition 1 can be proved by using  the following lemma.

{\bf Lemma.} {\em For all $m > 2$, $l > 2$ 
 \begin{equation}
  {\cal R}_{\pm}(m,l) \equiv  {\cal R}(x_{\pm}(m,l),m,l) < 0. 
 \label{r3.13.R}
 \end{equation}
}

 The proof of the Lemma is presented in the Appendix B. 

{\bf Proof of the Proposition 1.} Relation (\ref{r3.13.H})
follows from  (\ref{r3.8L}), ${\cal P}(x,m,l) = (l-1)(l-2) (x - x_{+})(x - x_{-})$ and Lemma.
Here 
 \begin{eqnarray}
  B_{\pm}(m,l) = \frac{{\cal R}_{\pm}(m,l)}{8 (l-1)^2 (l-2)^2 (x_{+} - x_{-})^{2} }  
           \nonumber  \\  
  = \frac{{\cal R}_{\pm}(m,l)}{32 (m-1) (l-1)(m + l - 3)}   < 0,  
  \label{r3.13.RB}
  \end{eqnarray} 
where  $m > 2$ and $l > 2$. Relation (\ref{r3.13.lim})  follows from (\ref{r3.13.H}) and (\ref{r3.13.RB}).
Thus, the Proposition 1 is proved. 

Now we analyze the behaviour of the function  $\lambda(x,m,l)$, for fixed 
$m,l$ and $x \neq x_{\pm}(m,l)$.
We find the  extremum  points  obeying $\frac{\partial}{\partial x} \lambda(x,m,l) = 0$.
By straightforward calculations we obtain 
\begin{eqnarray} 
  \frac{\partial}{\partial x} \lambda(x,m,l) = - f(x,m,l) ({\cal P}(x,m,l))^{-3},
                \qquad \label{r3.14a} \\
        f(x,m,l) = (l-1)(m-1)(x-1)(lx+m) \times
                 \nonumber \\
        \times [(l-2)x + m-1][(l-1)x + m-2],
                 \qquad \label{r3.14b}
\end{eqnarray}
$x \neq x_{\pm}(m,l)$. Using these relations we are led to the following
extremum points
 \begin{eqnarray} 
         x_a = 1,
            \qquad \label{r3.15a} \\
        x_b = x_b(m,l) \equiv - \frac{m-1}{l-2} < 0,
             \qquad \label{r3.15b} \\
        x_c = x_c(m,l) \equiv - \frac{m-2}{l-1} < 0,
                     \qquad \label{r3.15c} \\
        x_d = x_d(m,l) \equiv - \frac{m}{l} < 0.
                             \qquad \label{r3.15d}       
 \end{eqnarray}

These points obey  the following duality identities
 \begin{eqnarray} 
    x_d(m,l) = 1/x_d(l,m),
            \qquad \label{r3.16a} \\
     x_b(m,l) = 1/x_c(l,m).
             \qquad \label{r3.16b}       
  \end{eqnarray}
We also obtain the inequality
\begin{equation} 
    x_b(m,l) < x_c(m,l),
                 \qquad \label{r3.17a}
 \end{equation}
which is valid since 
\begin{equation} 
    x_c(m,l) - x_b(m,l) = \frac{m + l -3}{(l-1)(l-2)} > 0 
                 \qquad \label{r3.17b}
 \end{equation}
for all $m >2$, $l>2$.
   
   The points $ x_b, x_c, x_d$ obey the following 
   inclusion
  \begin{equation} 
   x_i(m,l) \in  (x_{-}(m,l),x_{+}(m,l)),    \qquad \label{r3.17x}
  \end{equation}
  $i = b,c,d$ for  $m >2$, $l>2$. 
  This inclusion just follows from relations  ${\cal P}_i(m,l) = {\cal P}(x_i(m,l),m,l) < 0$,
  $i = b,c,d$, since
  \begin{eqnarray} 
     {\cal P}_b(m,l) = - \frac{(m - 1)(m + l -3)}{(l - 2)} < 0,
              \qquad \label{r3.18a} \\
     {\cal P}_c(m,l) = - \frac{(m - 2)(m + l -3)}{(l - 1)} < 0,
               \qquad \label{r3.18b} \\
     {\cal P}_d(m,l) = - \frac{(l - 2) m^2 + 2 l m + (m - 2) l^2}{l^2} < 0,
                       \qquad \label{r3.18c}        
   \end{eqnarray}
 for all $m > 2$ and $l > 2$.

By using relations
\begin{eqnarray}
x_d - x_c &= \frac{m-2l}{l(l-1)}, 
\label{r3.19a} \\
x_d - x_b &= \frac{2m-l}{l(l-2)}, 
\label{r3.19b}
\end{eqnarray}
we get 
\begin{eqnarray}
(1) \ x_b < x_c < x_d, & \text{for } l < m/2,  \label{r3.20a} \\
(2) \ x_b < x_d < x_c, & \text{for }  m/2 < l < 2m, \label{r3.20b} \\
(3) \ x_d < x_b < x_c, & \text{for } l > 2m, \label{r3.20c},
\end{eqnarray}	
and 
\begin{eqnarray}
(1_0) \ x_b < x_c = x_d, & \text{if } l = m/2, \label{r3.21a} \\
(3_0) \ x_d = x_b < x_c, & \text{for } l = 2m. \label{r3.21b}
\end{eqnarray}	
We note that $m/2 < l < 2m$ may be rewritten as $l/2 < m < 2l$. 

Let us calculate $\lambda_i=\lambda(x_i,m,l)$, $i=a,b,c,d$.
We obtain
\begin{equation}
\lambda_a=-\frac{(m+l-1)(m+l)}{8(m+l-3)(m+l-2)}<0, \label{r3.22a}
\end{equation}

\begin{equation}
\lambda_b=\frac{lm^2+(l^2-8l+8)m+l^2-l}{8(l-2)(m-1)(l+m-3)}>0, \label{r3.22b}
\end{equation}

\begin{equation}
\lambda_c=\frac{ml^2+(m^2-8m+8)l+m^2-m}{8(m-2)(l-1)(l+m-3)}>0, \label{r3.22c}
\end{equation}
and 
\begin{equation}
\lambda_d=\frac{ml(m+l)}{8(lm^2+ml^2-2m^2-2l^2+2lm)}>0, \label{r3.22d}
\end{equation}
In the proof of inequalities  (\ref{r3.22b}), (\ref{r3.22c}) the following relations were
used: 
\begin{eqnarray}
   u(m,l) = l m^2+(l^2-8l+8)m+l^2-l  > 0,       \label{r3.23a} \\
   v(m,l) = m l^2+(m^2-8m+8)l+m^2-m  > 0        \label{r3.23b} \\
   w(m,l) = m^2(l-2) + l^2(m -2) + 2lm >0       \label{r3.23c}     
\end{eqnarray} 
for  $m > 2$, $l > 2$. Indeed,  for $m \ge 4$, $l \ge 4$  we get 
$u(m,l) = ml(m + l- 8) + 8m +l^2-l > 0$ and $u(4,3) = 26$, $u(3,4) = 24$, $u(3,3) = 12$.   
Thus, the first relation (\ref{r3.23a}) is valid. The second one (\ref{r3.23b})
just follows from the first one and $v(m,l)= u(l,m)$.

We also get 
\begin{equation}
\lambda_b-\lambda_c=\frac{(m-l)(m+l-3)}{4(l-2)(l-1)(m-2)(m-1)}  \begin{cases}
 >0, \text{if } m>l, \\
 =0, \text{if } m=l, \\
 <0, \text{if } m<l.
\end{cases}   \label{r3.24}
\end{equation}
and
\begin{eqnarray}
\lambda_d-\lambda_c=\frac{(m - 1)(m-2l)^3}{4(l-1)(m-2)(m+l-3)w(m,l)}, 
             \label{r3.25a} \\
\lambda_d-\lambda_b=\frac{(l-1)(l-2m)^3}{4(m-1)(l-2)(m+l-3)w(m,l)}, 
  \label{r3.25b}
\end{eqnarray}
  for  $m > 2$, $l > 2$. ( $w(m,l)$ is defined in (\ref{r3.23c}).)
By using these relations we obtain 
\begin{equation}
\lambda_d-\lambda_c \begin{cases}
 >0, \text{if } m > 2l, \\
 =0, \text{if } m = 2l, \\
 <0, \text{if } m < 2l.
\end{cases}   \label{r3.26}
\end{equation}
and 
\begin{equation}
\lambda_d - \lambda_b \begin{cases}
 >0, \text{if } l > 2m, \\
 =0, \text{if } l = 2m, \\
 <0, \text{if } l < 2m.
\end{cases}   \label{r3.27}
\end{equation}

Now we study the behaviour of the function $\lambda(x,m,l)$ with respect to variable $x$ for fixed  integer numbers $m >2$, $l>2$. We denote by $n(\Lambda, \alpha)$ the number of solutions 
(in $x$) of the equation  $\Lambda \alpha = \lambda(x,m,l)$. We calculate $n(\Lambda, \alpha)$
by using unequalities for points of extremum  $x_i$ and $\lambda_i$ ($i = b,c,d$) presented above and 
relations (\ref{r3.13.lim}) and  (\ref{r3.14b}).	

First, we start with the case $\alpha > 0$ and $x_{-} < x < x_{+}$. 
 
{\bf (1) $m > 2l$}. We get $x_b < x_c < x_d$ and  $\lambda_c < \lambda_d < \lambda_b$.
Here $x_b$ and $x_d$ are points of local maximum 
($x_b$ is a point of maximum on interval $(x_{-}, x_{+})$)
  and   $x_c$ is a point of local minimum.  We obtain          
 \begin{equation}
   n(\Lambda, \alpha) = \begin{cases}
   0, \  \Lambda \alpha > \lambda_b, \\
   1, \  \Lambda \alpha= \lambda_b, \\
   2, \  \lambda_d < \Lambda \alpha < \lambda_b, \\
   2, \   \Lambda \alpha = \lambda_d, \\
   4, \  \lambda_c < \Lambda \alpha < \lambda_d, \\
   3, \   \Lambda \alpha = \lambda_c, \\
   2, \   \Lambda \alpha < \lambda_c.
 \end{cases}   \label{r3.28}
 \end{equation}
Here and in what follows we use $x \neq x_d$.

We present an example of the function $\lambda(x) = \Lambda \alpha$ for $\alpha > 0$,  $m=12$ and $l=3$  at
Figure 1. At this and other figures  the point $(x_i, \lambda_i)$ is marked by $i$, where $i = a,b,c,d$.

\begin{figure}[!h]
	\begin{center}
		\includegraphics[width=0.75\linewidth]{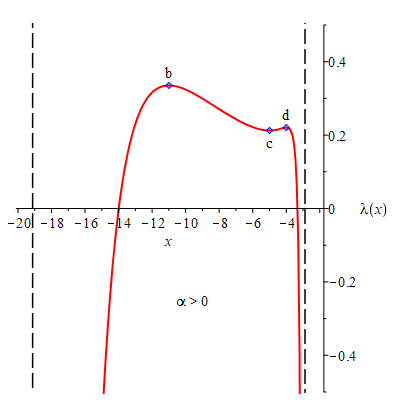}
		\caption{The function $\lambda(x) = \Lambda(x) \alpha$ for $\alpha > 0$,  $m=12$ and $l=3$.}
		\label{rfig:1}
	\end{center}
\end{figure}

 {\bf $(1_0)$ $m = 2l$}. We have $x_b < x_c = x_d$ and $\lambda_c = \lambda_d < \lambda_b$.
Here $x_b$ is the point of maximum on the interval $(x_{-}, x_{+})$ and   $x_c = x_d$
 is the point of inflection.
     We obtain          
  \begin{equation}
    n(\Lambda, \alpha) = \begin{cases}
    0, \  \Lambda \alpha > \lambda_b, \\
    1, \  \Lambda \alpha= \lambda_b, \\
    2, \  \lambda_d < \Lambda \alpha < \lambda_b, \\
     1, \   \Lambda \alpha = \lambda_d, \\
      2, \   \Lambda \alpha < \lambda_d. 
   \end{cases}   \label{r3.29}
  \end{equation}

An example of the function $\lambda(x) = \Lambda(x) \alpha$ for $\alpha > 0$,  $m=12$ and $l=6$ is depicted at
Figure 2.

\begin{figure}[!h]
	\begin{center}
		\includegraphics[width=0.75\linewidth]{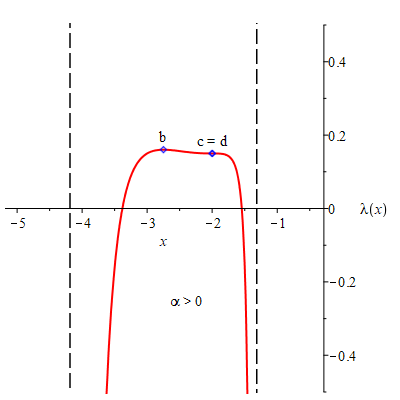}
		\caption{The function $\lambda(x) = \Lambda(x) \alpha$ for $\alpha > 0$,  $m=12$ and $l=6$.}
		\label{rfig:2}
	\end{center}
\end{figure}

 {\bf (2) $l/2 < m < 2l$.} 
 We obtain $x_b < x_d < x_c$ and $\lambda_d < \lambda_c$,  $\lambda_d  < \lambda_b$.
 The points $x_b,  x_c$ are points of local maximum and $x_d$ is point of local minimum.
  
  Now we split this case on three subcases: 
  $(2_{+})$ $l< m < 2l$, $(2_{0})$ $m = l$ and $(2_{-})$ $l/2 < m < l$.
 
 {\bf $(2_{+})$ $l < m < 2l$.} In this subcase we have $\lambda_d < \lambda_c < \lambda_b$ and 
 hence 

 \begin{equation}
   n(\Lambda, \alpha) = \begin{cases}
   0, \  \Lambda \alpha > \lambda_b, \\
   1, \  \Lambda \alpha= \lambda_b, \\
   2, \  \lambda_c < \Lambda \alpha < \lambda_b, \\
   3, \   \Lambda \alpha = \lambda_c, \\
   4, \  \lambda_d < \Lambda \alpha < \lambda_c, \\
   2, \   \Lambda \alpha = \lambda_d, \\
   2, \   \Lambda \alpha < \lambda_d.
 \end{cases}   \label{r3.30}
 \end{equation}
 
 An example of the function $\lambda(x) = \Lambda(x) \alpha$ for $\alpha > 0$,  $m=9$ and $l=7$ is depicted at
 Figure 3.
 
 \begin{figure}[!h]
 	\begin{center}
 		\includegraphics[width=0.75\linewidth]{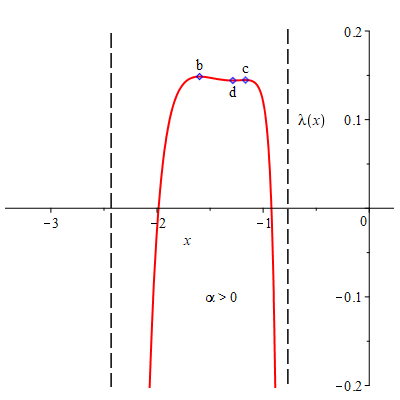}
 		\caption{The function $\lambda(x) = \Lambda(x) \alpha$ for $\alpha > 0$,  $m=9$ and $l=7$.}
 		\label{rfig:3}
 	\end{center}
 \end{figure}

{\bf $(2_{0})$ $m = l$.} In this subcase we find  $\lambda_d < \lambda_c = \lambda_b$.  
 Hence 

 \begin{equation}
   n(\Lambda, \alpha) = \begin{cases}
   0, \  \Lambda \alpha > \lambda_b = \lambda_c, \\
   2, \  \Lambda \alpha= \lambda_b = \lambda_c, \\
   4, \  \lambda_d < \Lambda \alpha < \lambda_b = \lambda_c, \\
   2, \   \Lambda \alpha = \lambda_d, \\
   2, \   \Lambda \alpha < \lambda_d.
 \end{cases}   \label{r3.31}
 \end{equation}

At  Figure 4 an example of the function $\lambda(x) = \Lambda(x) \alpha$ for $\alpha > 0$ and  $m=l=4$ is presented. 
 
 \begin{figure}[!h]
 	\begin{center}
 		\includegraphics[width=0.75\linewidth]{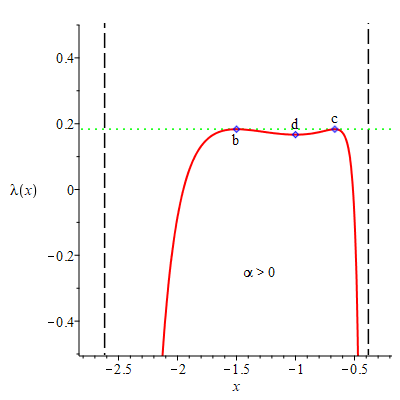}
 		\caption{The function $\lambda(x) = \Lambda(x) \alpha$ for $\alpha > 0$ and $m = l = 4$.}
 		\label{rfig:4}
 	\end{center}
 \end{figure}

 {\bf $(2_{-})$ $l/2 < m < l$.} In this subcase one gets  $\lambda_d < \lambda_b < \lambda_c$ and 
  \begin{equation}
    n(\Lambda, \alpha) = \begin{cases}
    0, \  \Lambda \alpha > \lambda_c, \\
    1, \  \Lambda \alpha= \lambda_c, \\
    2, \  \lambda_b < \Lambda \alpha < \lambda_c, \\
    3, \   \Lambda \alpha = \lambda_b, \\
    4, \  \lambda_d < \Lambda \alpha < \lambda_b, \\
    2, \   \Lambda \alpha = \lambda_d, \\
    2, \   \Lambda \alpha < \lambda_d.
  \end{cases}   \label{r3.32}
  \end{equation}

An example of the function $\lambda(x) = \Lambda(x) \alpha$ for $\alpha > 0$,  $m=4$ and $l=6$ is depicted at
 Figure 5.
 
 \begin{figure}[!h]
 	\begin{center}
 		\includegraphics[width=0.75\linewidth]{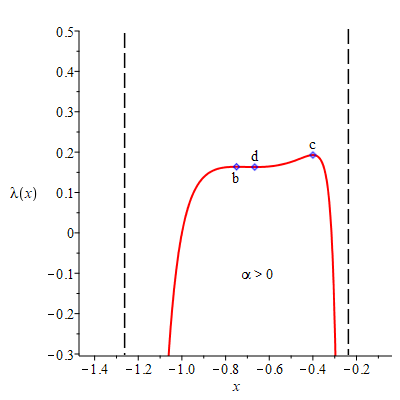}
 		\caption{The function $\lambda(x) = \Lambda(x) \alpha$ for $\alpha > 0$, $m =4$ and $l = 6$.}
 		\label{rfig:5}
 	\end{center}
 \end{figure}

{\bf (3) $m < l/2$}. We have $x_d < x_b < x_c$ and  $\lambda_b < \lambda_d < \lambda_c$. Here
$x_c$ and $x_d$ are points of local maximum ($x_c$ is the point of maximum on interval $(x_{-}, x_{+})$)
  and   $x_b$ is a point of local minimum.  We find           
 \begin{equation}
   n(\Lambda, \alpha) = \begin{cases}
   0, \  \Lambda \alpha > \lambda_c, \\
   1, \  \Lambda \alpha= \lambda_c, \\
   2, \  \lambda_d < \Lambda \alpha < \lambda_c, \\
   2, \   \Lambda \alpha = \lambda_d, \\
   4, \  \lambda_b < \Lambda \alpha < \lambda_d, \\
   3, \   \Lambda \alpha = \lambda_b, \\
   2, \   \Lambda \alpha < \lambda_b.
 \end{cases}   \label{r3.33}
 \end{equation}
 
 An example of the function $\lambda(x) = \Lambda(x) \alpha$ for $\alpha > 0$,  $m=4$ and $l=12$ is depicted at
  Figure 6.
  
  \begin{figure}[!h]
  	\begin{center}
  		\includegraphics[width=0.75\linewidth]{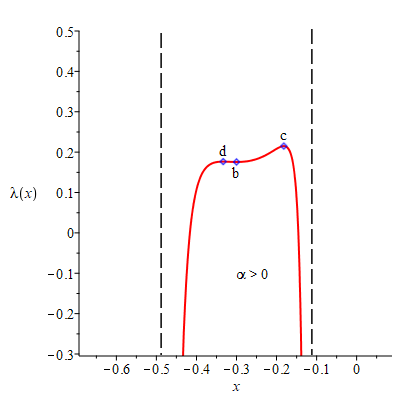}
  		\caption{The function $\lambda(x) = \Lambda(x) \alpha$ for $\alpha > 0$, 
  		$m =4$ and $l = 12$.}
  		\label{rfig:6}
  	\end{center}
  \end{figure}

{\bf $(3_0)$ $m = l/2$}. We get relations $x_d = x_b < x_c$ and  $\lambda_b = \lambda_d < \lambda_c$.
Here $x_c$ is the point of maximum on interval $(x_{-}, x_{+})$ and   $x_b = x_d$
 is the point of inflection.
     We obtain          
  \begin{equation}
    n(\Lambda, \alpha) = \begin{cases}
    0, \  \Lambda \alpha > \lambda_c, \\
    1, \  \Lambda \alpha= \lambda_c, \\
    2, \  \lambda_d < \Lambda \alpha < \lambda_c, \\
    1, \  \Lambda \alpha = \lambda_d, \\
    2, \   \Lambda \alpha < \lambda_d. \\
   \end{cases}   \label{r3.34}
  \end{equation}

{\bf Bounds on  $\Lambda \alpha$ for $\alpha > 0.$}
Summarizing all cases presented above we find that for $\alpha > 0$
exact solutions under consideration exist if and only if
 \begin{equation}
     \Lambda \alpha \leq \begin{cases}
     \    \lambda_b, {\rm for} \ m \geq l, \\
     \    \lambda_c,  {\rm for} \ m < l,
      \end{cases}   \label{r3.34s}
   \end{equation}
where $\lambda_b = \lambda_b (m,l)$ and  $\lambda_c = \lambda_c (m,l)$ 
are defined in  (\ref{r3.22b}) and  (\ref{r3.22c}), respectively.
For $m = 3$ and $l \geq 3$ we are led to relation  (\ref{0.i1}).

 An example of the function $\lambda(x) = \Lambda(x) \alpha$ for $\alpha > 0$,  $m=4$ and $l=8$ is presented at
  Figure 7. By using our analysis  we find  that for $\alpha > 0$ and small enough value of $\Lambda$
   there exist at least a pair of solutions $x_1, x_2$, obeying $x_{-} < x_1 < x_2 < x_{+} < 0$.
  
  \begin{figure}[!h]
  	\begin{center}
  		\includegraphics[width=0.75\linewidth]{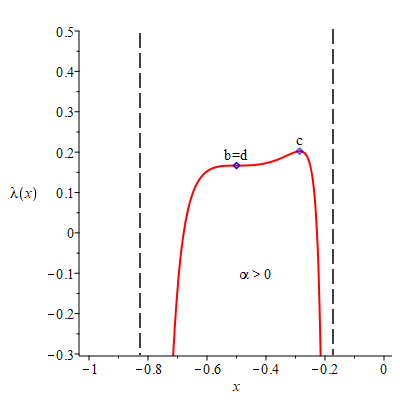}
  		\caption{The function $\lambda(x) = \Lambda(x) \alpha$ for $\alpha > 0$, $m =4$ and $l = 8$.}
  		\label{rfig:7}
  	\end{center}
  \end{figure}

Let us consider the case  $\alpha < 0$. 
We have $\Lambda |\alpha| = - \lambda(x) $, where $x < x_{-}$ or $x > x_{+}$.  Due to to 
the relations (\ref{r3.13.l}), (\ref{r3.14b})  and Proposition 1 
the function $- \lambda(x)$  is monotonically increasing in two intervals: i) in the
interval $(-\infty,x_{-})$ from $- \lambda_{\infty} $ to $+ \infty$ and ii) in the
interval $(x_{a} = 1, +\infty)$ from $- \lambda_{a} $ to $- \lambda_{\infty} $. 
The function $- \lambda(x)$  is monotonically decreasing in the  interval  
$(x_{+}, x_{a})$ from $+ \infty$ to $- \lambda_{a} $.
Here $x_{a}$ is a point of local minimum of the function $\Lambda(x) |\alpha| = - \lambda(x) $, which
is excluded from the solution and   $- \lambda_{a} < - \lambda_{\infty}$.  
This inequality may be readily verified: 
 due to relations (\ref{r3.13.l}), (\ref{r3.22a}) we obtain
  
 \begin{equation}
  \lambda_a - \lambda_{\infty} = \frac{2(m-1)[(2l-1)m + 2l(l-2)]}{8(l - 1)(l - 2)(m+l-3)(m+l-2)} > 0,
   \label{r3.35}
 \end{equation}
 for $m > 2$ and $l > 2$. The  functions 
 $\Lambda(x)  = \lambda(x)/\alpha$   for $\alpha = +1, -1$, respectively, and $m = l =4$ are 
 presented at Figure 8.

   For the number of solutions (for $\alpha < 0$) we obtain          
   \begin{equation}
     n(\Lambda, \alpha) = \begin{cases}
     2, \  \Lambda |\alpha| > |\lambda_{\infty}|, \\
     1, \  \Lambda |\alpha| = |\lambda_{\infty}|, \\
     2, \  |\lambda_{a}| < \Lambda |\alpha| < |\lambda_{\infty}|. \\
     0, \   \Lambda |\alpha| \leq |\lambda_{a}|. \\
    \end{cases}   \label{r3.36}
   \end{equation}
Here  $x \neq x_a =1$. Hence, for $\alpha < 0$ and big enough values of $\Lambda$
there exist  two solutions $x_1, x_2$: $x_1 < x_{-} < 0$  and  $x_2 > x_{+}$.

{\bf Bounds on  $\Lambda |\alpha|$ for $\alpha < 0.$}
It follows from (\ref{r3.36}) that for $\alpha < 0$
exact solutions  under consideration exist if and only if
 \begin{equation}
     \Lambda |\alpha| >     |\lambda_a| = \frac{(D -2)(D-1)}{8(D - 4)(D - 3)}, 
              \label{r3.36s}
   \end{equation}
where $\lambda_a $  is defined in  (\ref{r3.22a}).
This relation is valid for all $m > 2$, $l > 2$ ($D = m + l +1$), 
e.g. for $m = 3$ (see (\ref{0.i2})).

 \begin{figure}[!h]
   	\begin{center}
   		\includegraphics[width=0.80\linewidth]{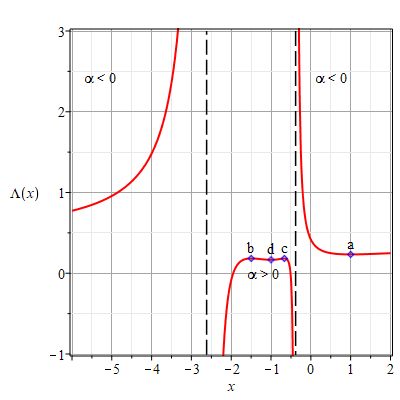}
   		\caption{The functions $\Lambda(x) = \lambda(x)/\alpha $ for $\alpha = \pm 1 $
   		and $m = l = 4$.}
   		\label{rfig:8}
   	\end{center}
   \end{figure}

\section{Stability analysis}

Here,  using results of refs. \cite{ErIvKob-16,Ivas-16}, we study the stability of exponential solutions (\ref{r2.2})
with non-static total volume factor, i.e. we put
\begin{equation}
  S_1(v) = \sum_{i = 1}^{n} v^i \neq 0.
  \label{r4.1}
\end{equation}

Here, as in  \cite{ErIvKob-16,Ivas-16}, we  impose the following restriction 
\begin{equation}
  ({\rm R }) \quad  \det (L_{ij}(v)) \neq 0
  \label{r4.2}
\end{equation}
on the symmetric matrix 
\begin{equation}
L =(L_{ij}(v)) = (2 G_{ij} - 4 \alpha G_{ijks} v^k v^s).
   \label{r4.1b}
 \end{equation}

For general cosmological ansatz with the (diagonal) metric 
\begin{equation}
 g= - dt \otimes dt + \sum_{i=1}^{n} e^{2\beta^i(t)}  dy^i \otimes dy^i,
 \label{r4.3}
\end{equation}
we have  the  set of (algebraic and differential) equations \cite{Iv-09,Iv-10}
\begin{eqnarray}
     E = G_{ij} h^i h^j + 2 \Lambda  - \alpha G_{ijkl} h^i h^j h^k h^l = 0,
         \label{r4.3.1} \\
         Y_i =  \frac{d L_i}{dt}  +  (\sum_{j=1}^n h^j) L_i -
                 \frac{2}{3} (G_{sj} h^s h^j -  4 \Lambda) = 0,
                     \label{r4.3.2a}
          \end{eqnarray}
where $h^i = \dot{\beta}^i$,           
 \begin{equation}
  L_i = L_i(h) = 2  G_{ij} h^j
       - \frac{4}{3} \alpha  G_{ijkl}  h^j h^k h^l  
       \label{r4.3.3},
 \end{equation}
 $i = 1,\ldots, n$.

Earlier, it was proved  \cite{Ivas-16} that a constant solution
$(h^i(t)) = (v^i)$ ($i = 1, \dots, n$; $n >3$) to eqs. (\ref{r4.3.1}), (\ref{r4.3.2a})
obeying restrictions (\ref{r4.1}), (\ref{r4.2}) is  stable under perturbations
\begin{equation}
 h^i(t) = v^i +  \delta h^i(t), 
\label{r4.3h}
\end{equation}
 $i = 1,\ldots, n$,  (as $t \to + \infty$)  in the following case
\begin{equation}
   S_1(v) = \sum_{k = 1}^{n} v^k > 0
 \label{r4.1a}
\end{equation}
and  it is unstable (as $t \to + \infty$) when 
\begin{equation}
   S_1(v) = \sum_{k = 1}^{n} v^k < 0.
 \label{r4.1c}
\end{equation}

For our considertion we have $ S_1(v) = m H + l h$ and hence due to $H >0$  the restriction
(\ref{r4.1a}) may be written in the following form
\begin{equation}
  x > - \frac{m}{l} = x_d,
 \label{r4.1x}
\end{equation}
while the  restriction (\ref{r4.1c}) may be written as
\begin{equation}
  x < - \frac{m}{l} = x_d.
 \label{r4.1y}
\end{equation}

The  perturbations $\delta h^i$ 
obey (in the linear approximation) the following set of linear equations
\cite{ErIvKob-16,Ivas-16}
 \begin{eqnarray}
   C_i(v) \delta h^i = 0, \label{r4.2C} \\
   L_{ij}(v) \delta \dot{h}^j =  B_{ij}(v) \delta h^j.
  \label{r4.3LB}
 \end{eqnarray}
  Here
 \begin{eqnarray}
 C_i(v)  =  2 v_{i} - 4 \alpha G_{ijks}  v^j v^k v^s, \label{r4.3.4} \\
 L_{ij}(v) = 2 G_{ij} - 4 \alpha G_{ijks} v^k v^s,
    \label{r4.3.5} \\
 B_{ij}(v) = - (\sum_{k = 1}^n v^k)  L_{ij}(v) - L_i(v) + \frac{4}{3} v_{j},
                     \label{r4.3.6}
 \end{eqnarray}
 where $v_i = G_{ij} v^j$,  $L_i(v) =  2 v_{i} - \frac{4}{3} \alpha  G_{ijks}  v^j v^k v^s$
 and $i,j,k,s = 1, \dots, n$.

In case when restrictions (\ref{r4.1}), (\ref{r4.2}) are imposed,  the set of  equations on  perturbations (\ref{r4.2C}), (\ref{r4.3LB})  has the following solution \cite{Ivas-16}
   \begin{eqnarray}
       \delta h^i = A^i \exp( -  S_1(v) t ),
       \label{r4.16}  \\
         \sum_{i =1}^{n} C_i(v)  A^i =0,
         \label{r4.16A}
   \end{eqnarray}
    ($A^i$ are constants) $i = 1, \dots, n$.    

It was shown in  \cite{Ivas-16} that  for  the vector $v$ from  (\ref{r3.1}), obeying
 relations (\ref{r3.3}), the matrix $L$ is a block-diagonal one
\begin{equation}
 (L_{ij}) = {\rm diag}(L_{\mu \nu}, L_{\alpha \beta} ),
 \label{r4.5}
\end{equation}
where
\begin{eqnarray}
  L_{\mu \nu} =  G_{\mu \nu} (2 + 4 \alpha S_{HH}),
 \label{r4.6HH}     \\
  L_{\alpha \beta} = G_{\alpha \beta} (2 + 4 \alpha S_{hh}) 
  \label{r4.6hh}
\end{eqnarray}
and
\begin{eqnarray}
  S_{HH} =  (m-2)(m -3) H^2  + 2(m-2)lHh + l(l - 1)h^2 ,
  \label{r4.7}   \\
  S_{hh} = m(m-1)H^2  + 2m(l - 2)Hh+ (l- 2)(l- 3)h^2.
  \label{r4.8}
\end{eqnarray}

The matrix (\ref{r4.5}) is invertible  only if  $m > 1$,  $l > 1$ and
\begin{eqnarray}
 S_{HH} \neq - \frac{1}{2 \alpha}, \label{r4.9a}
  \\
 S_{hh} \neq - \frac{1}{2 \alpha}.
 \label{r4.9b}
\end{eqnarray}
We remind (the reader) that  the matrices  $(G_{\mu \nu}) = (\delta_{\mu \nu} -1 )$ and
$(G_{\alpha \beta}) = (\delta_{\alpha \beta} - 1)$ are invertible only if  $m > 1$ and $l > 1$.

Now, we prove that inequalities (\ref{r4.9a}), (\ref{r4.9b}) are obeyed
if   
 \begin{equation}
  x \neq  -  \frac{m-2}{l -1} = x_c
 \label{r4.12a} 
 \end{equation}
and
 \begin{equation}
     x \neq  - \frac{m -1}{l - 2} = x_b 
 \label{r4.12b} 
 \end{equation}
for $ l > 2$.

Let us suppose that  (\ref{r4.9a}) does not take place, i.e.  
$S_{HH} = - \frac{1}{2 \alpha}$.  Then using (\ref{r3.5}) we obtain  
\begin{equation}
S_{HH} - Q = - 2 (H - h) ((m-2) H + (l -1) h ) = 0, 
\label{r4.10a}                                            
\end{equation}
which implies  due to $H - h \neq 0$ (see  (\ref{r3.3}))         
\begin{equation}
   (m-2) H + (l -1) h = 0.
\label{r4.11a}                                                        
\end{equation}
 This relation contradicts to the  restriction (\ref{r4.12a}). 
 The obtained contradiction proves the inequality (\ref{r4.9a}).     
      
 Now let us  suppose that  (\ref{r4.9b}) is not valid,
 i.e. $S_{hh} = - \frac{1}{2 \alpha}$. Then using (\ref{r3.5}) we find  
 \begin{equation}
 S_{hh} - Q =  - 2 (h - H)  ((l - 2) h + (m -1) H ) = 0. 
 \label{r4.10b}                                            
 \end{equation}
 Due to $H - h \neq 0$ this implies           
 \begin{equation}
    (l - 2) h + (m -1) H = 0,
 \label{r4.11b}                                                        
 \end{equation}
  which is in the contradiction with the restrictions (\ref{r4.12b})
  and $H > 0$. This contradiction lead  us to the proof of the inequality
 (\ref{r4.9b}).         

Thus, we have proved that relations (\ref{r4.9a}) and (\ref{r4.9b}) are valid and hence the
 restriction (\ref{r4.2}) is satisfied for our solutions. 

Thus we have proved the following proposition. 

{\bf Proposition 2.} {\em The cosmological solutions under consideration, which obey  $x = h/H \neq x_i$, $i = a,b,c,d$, where $x_a =1$, $x_b = - \frac{m -1}{l - 2}$, $x_c = - \frac{m-2}{l -1}$,  $x_d = - \frac{m}{l}$, 
are  stable if i) $x > x_d$ and unstable if ii) $x < x_d$. }

Here it should be noted that our anisotropic solutions with non-static volume factor
are not defined for $x= x_a$ and $x= x_d$. Meanwhile, they
are defined when $x = x_b$ or $x = x_c$, if $x \neq x_d$. The stability 
analysis of these special solutions can not be covered by the equations for perturbations 
(\ref{r4.2C}), (\ref{r4.3LB}) in the linear approximation. As it was pointed 
 out in ref. \cite{ChTop-17} this analysis needs a special consideration.  

Now we  consider  the number of non-special stable solutions which are
given by Proposition 2 (see item i)). We denote this number as $n_{+}(\Lambda, \alpha)$.
By using the results  from the previous section (e.g. illustrated by figures) we 
obtain for $\alpha > 0$:   

 { \bf (1), $(1_{0})$ $m \geq 2l$}          
  \begin{equation}
    n_{+}(\Lambda, \alpha) = \begin{cases}
    0, \  \Lambda \alpha \geq \lambda_d, \\
    1, \  \Lambda \alpha < \lambda_d; \\
    \end{cases}   \label{r4.13c}
  \end{equation}
 
  {\bf (2) $l/2 < m < 2l$} 
     \begin{equation}
    n_{+}(\Lambda, \alpha) = \begin{cases}
    0, \  \Lambda \alpha \geq \lambda_c, \\
    2, \  \lambda_d < \Lambda \alpha < \lambda_c, \\
    1, \   \Lambda \alpha \leq \lambda_d; \\
   \end{cases}   \label{r4.14c}
  \end{equation}
  
  {\bf (3) $ m < l/2$}         
  \begin{equation}
    n_{+}(\Lambda, \alpha) = \begin{cases}
    0, \  \Lambda \alpha \geq \lambda_c, \\
    2, \  \lambda_d \leq \Lambda \alpha < \lambda_c, \\
    3, \  \lambda_b < \Lambda \alpha < \lambda_d, \\
    1, \   \Lambda \alpha \leq \lambda_b;
  \end{cases}   \label{r4.15c}
  \end{equation}
  
   {\bf $(3_0)$ $m = l/2$}          
   \begin{equation}
     n_{+}(\Lambda, \alpha) = \begin{cases}
     0, \  \Lambda \alpha \geq \lambda_c, \\
     2, \ \lambda_d < \Lambda \alpha < \lambda_c, \\
     1, \  \Lambda \alpha \leq \lambda_d. \\
    \end{cases}   \label{r4.16c}
   \end{equation}
 
 We see, that for $\alpha > 0$ and small enough value of $\Lambda$ there exists at least one stable solution 
 with $x \in (x_{-},x_{+})$. 
 
 {\bf Bounds on  $\Lambda \alpha$ for stable solutions with $\alpha > 0.$}
 Summarizing all cases presented above we find that for $\alpha > 0$
 stable exact solutions under consideration exist if and only if
  \begin{equation}
      \Lambda \alpha < \begin{cases}
      \    \lambda_d, {\rm for} \ m \geq 2l, \\
      \    \lambda_c,  {\rm for} \ m < 2l,
       \end{cases}   \label{r3.34s}
    \end{equation}
 where $\lambda_c = \lambda_c (m,l)$ and  $\lambda_d = \lambda_d (m,l)$ 
 are defined in  (\ref{r3.22c}) and  (\ref{r3.22d}), respectively.
 For $m = 3$ and $l > 2$ we are led to relation  
  $\Lambda \alpha <  \lambda_c$ instead of  (\ref{0.i1}).

 In the case  $\alpha < 0$ we obtain          
    \begin{equation}
      n_{+}(\Lambda, \alpha) = \begin{cases}
      1, \  \Lambda |\alpha| \geq |\lambda_{\infty}|, \\
     
      2, \  |\lambda_{a}| < \Lambda |\alpha| < |\lambda_{\infty}|, \\
      0, \   \Lambda |\alpha| \leq |\lambda_{a}|. \\
     \end{cases}   \label{r4.17}
    \end{equation}
 Here the inequality  $x \neq x_a =1$ was used. Our analysis tells us that 
  for $\alpha < 0$ and big enough value of $\Lambda$
 there exists at least one stable solution governed by $x$ which obey $x > x_{+}$. 
 We also find that the solution with $x < x_{-}$ is  unstable. 
 
 {\bf Bounds on  $\Lambda |\alpha|$ for stable solutions with $\alpha < 0.$}
 It follows from (\ref{r4.17}) that for $\alpha < 0$
 stable exact solutions  under consideration exist if and only if
 the relation (\ref{r3.36s}) ($\Lambda |\alpha| >   |\lambda_a|$)  is obeyed.

\section{Solutions describing a small enough variation of $G$}

  Here we analyze the solutions  by using the restriction on variation of the effective
  gravitational constant $G$, which is inversely proportional (in the Jordan frame) 
  to the  volume scale factor  of the (anisotropic) internal space \cite{IvKob,BIM,Mel} 
  (see also references therein), i.e.

  \begin{equation}
  \label{r5.G0}
    G = {\rm const } \exp{[- (m - 3) H t - l h t]}.
   \end{equation}
 
By using (\ref{r5.G0}) one can get the following formula 
for a dimensionless parameter of temporal variation of $G$ ($G$-dot):
\begin{equation}
  \delta \equiv \frac{\dot{G}}{GH}  =  - (m - 3 + l x), \qquad x = h/H.
\label{r5.G}
\end{equation}
Here  $H >0$ is the Hubble parameter.

Due to observational data, the variation of the gravitational constant is 
on the level of $10^{-13}$ per year and less.
For example, one can use, as it was done in ref. \cite{IvKob},  the following bounds on the 
value of the dimensionless variation of the  effective gravitational constant:

 \begin{equation}
 \label{r5.G1}
  - 0,65 \cdot 10^{-3} < \delta < 1,12 \cdot 10^{-3}.
 \end{equation}
They come from the most stringent limitation
on $G$-dot obtained by the set of ephemerides \cite{Pitjeva}
and  value of the Hubble parameter (at present) \cite{Ade}
 when both are written with 95\% confidence level \cite{IvKob}.

  When the value  $\delta$ is fixed we get from (\ref{r5.G})
    \begin{equation}
  x =  x_0(\delta) = x_0(\delta,m,l) \equiv  - \frac{(m - 3 + \delta)}{l}. 
  \label{r5.0}    
 \end{equation}

We remind (the reader) that our solutions are defined if 
 \begin{equation}
      {\cal P}(x_0(\delta,m,l),m,l) \neq 0, 
      \label{r5.P}    
     \end{equation}
 or, if
 \begin{equation}
   x_0(\delta,m,l) \neq  x_{\pm}(m,l). 
  \label{r5.x}    
 \end{equation}

The substitution of  $ x = x_0(\delta,m,l)$ into quadratic polynomial (\ref{r3.7}) 
gives us 
\begin{eqnarray}
 {\cal P}(x_0(\delta,m,l),m,l)  = {\cal P}(x_0(0,m,l),m,l)
    \nonumber    \\
   -  4 \frac{ (l-1)(m + l -3) }{l^2} \delta  +  \frac{(l-1)(l-2) }{l^2} \delta^2, 
     \label{r5.1}
 \end{eqnarray}
where (see  \cite{ErIv-17-1})
\begin{equation}
  {\cal P}(x_0(0,m,l),m,l) \equiv  {\cal P}_0(m,l) =   \frac{1}{l^2} (m + l - 3)[(5-m)l +2m -6]. 
     \label{r5.2}
 \end{equation}
 
 We note that equation ${\cal P}_0(m,l) = 0$ implies relation 
 $l = l_0(m) = \frac{2m - 6}{m - 5} = 2 + \frac{4}{m - 5}$, $m \neq 5$. 
 For $m > 9$ we get $2 < l_0(m) < 3$, that
 means that integer solutions are absent in this interval. 
 For $3 \leq m \leq 9$ and  $m \neq 5$, the only integer 
 values of $l_0(m) > 2$  takes place for $m = 6,7,9$ and we get a special set of pairs $(m,l)$:
   \begin{equation}
     A) \quad (m,l) = (6,6), (7,4), (9,3), 
        \label{r5.3c}
    \end{equation}
 which was obtained in \cite{ErIv-17-1}. In the case A) the restriction (\ref{r5.P})
 gives us (see (\ref{r5.1})) $\delta \neq 0$ and $- 4 (l +m -3) \delta + (l-2) \delta^2 \neq 0$ 
 for $l > 2$ which lead
 us to two restrictions: $\delta \neq 0$ and $\delta \neq  4\frac{l + m -3}{l-2} = 9, 16,36$ for 
 $(m,l) = (6,6), (7,4), (9,3)$, respectively.
 But the second one may be omitted due to bounds (\ref{r5.G1}). 

    Let us consider the second case
   \begin{equation}
           B) \quad m =5, \quad l > 2. 
              \label{r5.4}
   \end{equation}
     In this case the restriction (\ref{r5.P}) reads
     \begin{equation}
      4 (l + 2) - 4 (l-1) (l+ 2) \delta  +  (l-1)(l-2)  \delta^2 \neq 0,
      \label{r5.5}    
      \end{equation}
   $l > 2$.  It may be rewritten
   as  
   \begin{equation}
      \delta  \neq  \delta_{\pm}(5,l) \equiv  2 \frac{(l+2)}{(l-2)}
       \left(1 \pm \sqrt{\frac{l^2}{(l-1)(l + 2)}} \right).
         \label{r5.6}
    \end{equation}
  The first restriction  $\delta  \neq  \delta_{+}(5,l)$, $l >2$, may be omitted due to the bounds
  (\ref{r5.G1}) since $\delta_{+}(5,l) > 2 \frac{(l + 2)}{(l-2)} >  2$ for $l > 2$. So,
  the only  second restriction $\delta  \neq  \delta_{-}(5,l)$, $l >2$, should be imposed.
  Since 
  \begin{equation}
  \delta_{-}(5,l) =  \frac{2}{l-1} \left(1 + \sqrt{\frac{l^2}{(l-1)(l + 2)}} \right)^{-1} \sim 1/(l-1)
   \label{r5.6a}
  \end{equation}
  as $l \to + \infty$, this restriction forbids one 
  value  of $\delta$
  obeying the bounds  (\ref{r5.G1}) for big anough value of $l$ (e.g., for $ l > 1000$). 
      
Now we consider the last case
  \begin{equation}
   C) \ (m,l)  {\rm \ do \ not \ belong \ to \ cases \ A \ and \ B }. 
    \label{r5.7}
  \end{equation}

In the case C) the restriction (\ref{r5.P}) reads
     \begin{equation}
       (l + m - 3) [(5-m)l +2m -6] - 4 (l-1) (l + m - 3) \delta  +  (l-1)(l-2)  \delta^2 \neq 0,
      \label{r5.8}    
     \end{equation}
   $l > 2$.  It may be rewritten
   as  
   \begin{equation}
      \delta  \neq  \delta_{\pm}(m,l) \equiv  2 \frac{(l + m - 3)}{(l-2)}
       \left(1 \pm \sqrt{\frac{l^2 (m-1)}{4(l-1)(l + m - 3)}} \right).
         \label{r5.9}
    \end{equation}
  The first restriction  $\delta  \neq  \delta_{+}(m,l)$ ($l >2$) may be omitted due to the bounds
  (\ref{r5.G1}) since $\delta_{+}(m,l) > 2 \frac{(l + m -3)}{(l-2)} >  2$ for $m >2$, $l > 2$. So,
  the only  second restriction $\delta  \neq  \delta_{-}(m,l)$, should be imposed.
  Here another equivalent relation may be used
  \begin{equation}
  \delta_{-}(m,l) =  \frac{(5-m)l +2m -6}{2(l-1)}
   \left(1 + \sqrt{\frac{l^2 (m-1) }{4(l-1)(l + m - 3)}} \right)^{-1}. 
   \label{r5.9d}
  \end{equation}

Thus, for our special values of $\delta$ obeying the bounds  (\ref{r5.G1}) the
only restriction on $\delta$ coming from $(\ref{r5.x})$ or $(\ref{r5.P})$ are the following
ones 
 \begin{eqnarray}
      \delta \neq 0, {\rm \ in \ the \ case \ A},  \label{r5.9a}\\ 
      \delta  \neq  \delta_{-}(5,l),  {\rm \ in \ the \ case \ B} \label{r5.9b},\\     
      \delta  \neq  \delta_{-}(m,l), {\rm \ in \ the \ case \ C} \label{r5.9c}, 
 \end{eqnarray}
where $\delta_{-}(5,l)$ is defined in (\ref{r5.6a}) and $\delta_{-}(m,l)$ is defined in (\ref{r5.9d}).

Now we analyse the stability of these special solutions. The main condition
for stability $x_0(\delta) > x_d$ is satisfied  since
 \begin{equation}
   x_0(\delta) -  x_d = \frac{3 - \delta}{l} > 0 
  \label{r5.10}
  \end{equation}
 due to our bounds  (\ref{r5.G1}).
 
 Other three conditions (see Proposition 2) : 
 $x_0(\delta) \neq  x_a$, $x_0(\delta) \neq  x_b$ and $x_0(\delta) \neq  x_c$ read  
   \begin{eqnarray}
        \delta \neq \delta_a =  - (m + l - 3),    \label{r5.11a}\\
        \delta \neq \delta_b = \frac{2(m + l - 3)}{l - 2},  \label{r5.11b}\\ 
        \delta  \neq  \delta_c = \frac{2(m + l - 3)}{l - 1} \label{r5.11c}.     
   \end{eqnarray}
They are satisfied due to bounds (\ref{r5.G1}) and inequalities:  
$\delta_a \leq  -3$, $\delta_b > 2$ and $\delta_c > 1$.  

Thus, we have shown that all well-defined solutions under consideration, which  obey restrictions 
(\ref{r5.9a}), (\ref{r5.9b}), (\ref{r5.9c}) and the physical bounds (\ref{r5.G1}), are stable.   

\section{Conclusions}

We have considered the  $D$-dimensional  Einstein-Gauss-Bonnet (EGB) model
with the $\Lambda$-term and two non-zero constants $\alpha_1$ and $\alpha_2$.  
By dealing with diagonal  cosmological  metrics, we have found 
for certain  (fine-tuned) $\Lambda = \Lambda(x,m,l,\alpha)$ with  
$\alpha = \alpha_2 / \alpha_1 $ a class of solutions with  exponential 
time dependence of two scale factors.
This exponential dependence is governed by two Hubble-like parameters $H >0$ and $h$, 
corresponding to submanifolds of dimensions $m > 2$ and $l > 2$, respectively, with  $D = 1 + m + l$. 
Here $m > 2$ is the dimension  of the expanding subspace and $l > 2$
is the dimension of another  one, the dimensionless parameter $x = h/H$ satisfies the 
following restrictions:  $x \neq 1$,  $x \neq x_d = - m/l$ and 
 $(m - 1)(m - 2) + 2 (m - 1)(l - 1) x   + (l - 1)(l - 2)x^2 \neq 0$.

Any obtained solution describes an exponential expansion of  $3$-dimensional subspace
 (which may be identified with ``our'' space) with
the Hubble parameter $H > 0$ and anisotropic behaviour of $(m-3+ l)$-dimensional internal space:
expanding in $(m-3)$ dimensions (with Hubble-like parameter $H$) and  either contracting, 
or expanding (with Hubble-like parameter $h$) or stable in $l$ dimensions.
The solutions are governed by master equation $\Lambda(x,m,l,\alpha) = \Lambda$, which 
may be solved in radicals for all values of $\Lambda$, since it is equivalent to a polynomial equation of either  
fourth or third order (depending upon $\Lambda$). The analytical solution for $m = l$ \cite{IvKob-18mm} 
 is presented in  Appendix A.

Here we have obtained the bounds on  $\Lambda$ which guarantee the existence
of the exponential cosmological solutions under consideration:
 \begin{equation}
     \Lambda \alpha \leq \lambda_{*}
      \label{c.1}
   \end{equation}
 for $\alpha > 0$ and
 \begin{equation}
     \Lambda |\alpha| >    \frac{(D -2)(D-1)}{8(D - 4)(D - 3)} 
              \label{c.2}
   \end{equation}
for $\alpha < 0$. In (\ref{c.1}) we denote: $\lambda_{*} = \lambda_b$ for $m \geq l$ 
and  $\lambda_{*} = \lambda_c$   for $m < l$, where  $\lambda_b$  and  $\lambda_c$ 
are defined in (\ref{r3.22b})  and  (\ref{r3.22c}), respectively. These bounds 
generalize the bounds (\ref{0.i1}) and (\ref{0.i2}) for $m =3$ \cite{Pavl-p-18}.
It should be noted that the bounds (\ref{c.1}) and (\ref{c.2}) were obtained here  without solving 
the equations of motion (e.g. the master equation). They were obtained by 
analyzing the function $\lambda = \lambda(x,m,l)$      from (\ref{r3.8L}) ($\lambda =  \Lambda \alpha$ ), 
e.g. by using the ``duality'' identity  $\lambda(x,m,l) = \lambda (1/x,l,m)$. 
(The ``duality'' transformation $(x,m,l) \mapsto (1/x,l,m)$ describes  just a trivial 
 interchange of factor spaces which corresponds to the replacement $(H,h, m,l) \mapsto (h, H,l,m)$.)
  
Using the scheme, which was developed in ref. \cite{Ivas-16},  
we  have proved that any of these solutions obeying the additional restrictions: 
$x \neq  - \frac{m -2}{l - 1}$ and  $x \neq  - \frac{m -1}{l - 2}$,
 is stable (as  $t \to + \infty$) if  $x > x_d = - m/l$ and unstable if  $x < x_d$.

We have also found that for $\alpha > 0$  stable exact solutions
exist if and only if: 
\begin{equation}
 \Lambda \alpha < \lambda_d  \label{c.3}
 \end{equation}
 for $m \geq 2l$    and  
 \begin{equation}
 \Lambda \alpha <  \lambda_c,  \label{c.4}
 \end{equation}
 for $m < 2l$,     
 where   $\lambda_d$ is defined in   (\ref{r3.22d}).
 For $\alpha < 0$  stable exact solutions exist only if
 the relation (\ref{c.2})   is obeyed.

It was also shown that all (well-defined) solutions with small enough varation of 
the effective gravitational constant $G$ (in the Jordan frame)
are stable.

Here an open problem is to extend  the cosmological solutions from this paper 
to the cosmological type solutions in the Lovelock gravitational model \cite{Lov},
 e.g. to solutions  describing cosmological and static configurations. Another problem
is related to  search and analysis of  the solutions with three factor spaces.
These and some orther topics may be addressed in our separate publications. 

It should be noted here that the results obtained in this paper and its possible  extensions to static and other
 cases may  be used in  other areas of physics (e.g. chromodynamics, condensed matter etc) by  applying powerful holographic methods based on  $AdS/CFT$, $dS/CFT$ approaches and its generalizations. 


\renewcommand{\theequation}{\Alph{subsection}.\arabic{equation}}
\renewcommand{\thesection}{}
\renewcommand{\thesubsection}{\Alph{subsection}}
\setcounter{section}{0}

\section{Appendix}

\subsection{The analytical solution for $m=l$}

For any $m = l > 2$ the master equation (\ref{r3.13.M})  reads
 \begin{equation}
   A x^4+B x^3 + C x^2 + B x + A = 0,      
               \qquad    \label{3.13.Mx} 
 \end{equation} 
where
\begin{eqnarray} 
A= 8  \lambda (m-2)^2(m-1) + m(m+1)(m-2), \label{3.13.A}    \\
B= 32 \lambda (m-2)(m-1)^2 + 4 m (m-1)^2, \label{3.13.B} \\
C= 16 \lambda (m-1)(3m^2-8m+6) + 2m(m-1)(3m -4). \label{3.13.C}
\end{eqnarray} 

It may be readily solved  in radicals, by using the substitution $y = x + \frac{1}{x}$ \cite{IvKob-18mm}.
For $A \neq 0$ we obtain 

\begin{equation}
\begin{aligned}
x= \frac{1}{4A} \left( - B + \nu_1 \sqrt{E - 2 B \nu_2 \sqrt{d}} + \nu_2 \sqrt{d} \right), 
 \label{3.13.x}
\end{aligned}
\end{equation}
where $\nu_1 = \pm 1$, $\nu_2 = \pm 1$ and 
\begin{equation} 
 d = 8A^2 - 4 C A + B^2, \qquad  E = - 8A^2 - 4CA + 2 B^2.
             \label{3.13.E}
 \end{equation}
We get 
\begin{eqnarray} 
d  = 16 m^2(2m^2-7m+7)  
- 128 m(m-1)(m-2) (2m-3) \lambda   
 \label{3.13.dd}, \\
 E = 1024 \lambda ^2(m-2)^2(m-1)^2(2m-3)   \nonumber \\
        + 128 \lambda (m-2)(m-1)m(4m-7)  \nonumber \\
        -16 m^2 (2 m^3- 11 m^2+15 m-4). 
             \label{3.13.EE}
 \end{eqnarray}

 For $A = 0$,   the  solution reads 
 \begin{equation}
 \begin{aligned}
 x= \frac{1}{2B} [ - C \pm \sqrt{C^2 -4 B^2 } ], \  {\rm or} \ x=0,
  \label{3.13.x0}
 \end{aligned}
 \end{equation}
where 
\begin{equation}
\begin{aligned}
 B = - 8 m (m-1), \qquad C = - \frac{4m}{m-2} (4 m^2-10m+7). 
 \label{3.13.BC}
\end{aligned}
\end{equation}

The special solution for $m=3$ was considered recently in ref. \cite{IvKob-m3-18}.

\subsection{The proof of the Lemma}

Here we give the proof of the Lemma from Section 2.
The calculations (by using Mathematica) lead us to following relations
\begin{equation}
{\cal R}_{\pm}(m,l)={\cal R}(x_{\pm}(m,l),m,l)=\frac{A(m,l) \pm B(m,l)\sqrt{\Delta(m,l)}}{C(l)} \label{rA.1}
\end{equation}
where
\begin{align*}
A(m,l)&=-2(m-1)(l+m-3)A_{*}(m,l),\\
A_{*}(m,l)&=l^2m^2+4lm^2-4m^2+l^3m-4l^2m-8lm+8m-2l^3+8l^2-4l,\\
B(m,l)&=8l(m-1)^2(l+m-3)>0,\\
\Delta(m,l)&=(m-1)(l-1)(l+m-3)>0,\\
C(l)&=  (l - 2)^3 (l -1) > 0.
\end{align*}

In order to prove ${\cal R}_{-}(m,l) <0$  it is sufficient to prove that $A_{*}(m,l) > 0$ 
for $m>2$ and $l>2$.

Let $m \ge 4$. Then we group $A_*(m,l)$ as the sum of the non-negative terms:

\begin{eqnarray}
 &A_*(m,l)= (l^2m^2-4l^2m)_1+(4lm^2-4m^2-8lm)_2 \nonumber \\
            &+(l^3m-2l^3)_3+(8m)_4+(8l^2-4l)_5; \nonumber 
\end{eqnarray}
where
\begin{align*}
(.)_1&=l^2m^2-4l^2m=l^2m(m-4)\ge0, \\
(.)_2&=4lm^2-4m^2-8lm=2(l-2)m^2+2lm(m-4)>0, \\
(.)_3&=l^3(m-2)>0,  \\
(.)_4&>0, \\
(.)_5&=4l(2l-1)>0. 
\end{align*}

Thus, we get  $A_{*}(m,l) > 0$ for $m \ge 4$ and $l > 2$.
For $m=3$ we have $A_*(3,l)=l^3+5l^2+8l-12 \ge 84 $ (as $l\ge3$).
Thus,  ${\cal R}_{-}(m,l) < 0$  ($m>2$, $l>2$) is proved.

Now we prove ${\cal R}_{+}(m,l) < 0$ ($m>2$, $l>2$)).
By using the identities  (\ref{r3.9dR}), (\ref{r3.12}) 
and definitions of ${\cal R}_{\pm}(m,l)$ we obtain
\begin{eqnarray}
{\cal R}_{+}(m,l)= {\cal R}(x_{+}(m,l),m,l) 
= (x_{+}(m,l))^4 {\cal R}(\frac{1}{x_{+}(m,l)},l,m) \nonumber \\
= (x_{+}(m,l))^4 {\cal R}(x_{-}(l,m),l,m) = 
   (x_{+}(m,l))^4 {\cal R}_{-}(l,m) < 0.
                               \label{rA.2}
\end{eqnarray}
By this we complete the proof of the Lemma. 

\vspace{0.2truecm}

 {\bf Acknowledgments}

The publication has been prepared with the support of the ``RUDN University Program 5-100''.
It was also partially supported by the  Russian Foundation for Basic Research,  grant  Nr. 16-02-00602.


\end{document}